\documentclass[aps,prl, notitlepage, amsmath,amssymb,nofootinbib,showpacs,superscriptaddress]{revtex4-1}
\usepackage[english]{babel}
\usepackage{latexsym}
\usepackage{graphics}
\usepackage{graphicx}
\usepackage{epsfig}
\usepackage{color}
\usepackage{bm}
\usepackage{amsmath}
\usepackage{amssymb}
\usepackage{amsthm}
\usepackage{dcolumn}
\usepackage{bm}
\usepackage{float}
\usepackage{hyperref}
\usepackage{color}
\usepackage{epstopdf}
\usepackage{cleveref}
\usepackage{ulem}
\usepackage[svgnames]{xcolor}
\hypersetup{hidelinks,colorlinks=true,allcolors=DarkBlue}

\usepackage{array}
\usepackage{makecell}
\usepackage{braket}
\usepackage{graphicx}

\theoremstyle{remark}

\begin{document}

\title{High-performance state-vector emulator \\ of a gate-based quantum processor implemented in the Rust programming language}

\author{Ilya A. Luchnikov}
\email{luchnikovilya@gmail.com}
\affiliation{Russian Quantum Center, Skolkovo, Moscow 143025, Russia}

\author{Oleg E. Tatarkin}
\affiliation{Russian Quantum Center, Skolkovo, Moscow 143025, Russia}

\author{Aleksey K. Fedorov}
\email{akf@rqc.ru}
\affiliation{Russian Quantum Center, Skolkovo, Moscow 143025, Russia}

\begin{abstract}
We propose a high-performance state-vector emulator of a gate-based quantum processors developed in the Rust programming language. It supports OpenQASM 2.0 programming language \cite{cross2017open} for quantum circuits specification and has a user-friendly Python based API. We present a wide range of numerical benchmarks of the emulator. We expect that our emulator will be used for design and validation of new quantum algorithms.
\end{abstract}

\maketitle

\begin{widetext}

\section{Introduction}

Quantum computing is believed to be useful for solving various classes of economically impactful problems~\cite{Fedorov2022}. Currently, available quantum information processing devices are able to solve certain computational problems that are close to the limits of what can be done with the most powerful classical technologies~\cite{Fedorov2022,Martinis2019,Pan2020}. 
However, quantum devices are still very expensive and represented by a few copies all over the world. Therefore, they still do not suit as a platform for massive development and testing of new quantum algorithms by a huge community of researchers.

This problem is solved by quantum emulators~\cite{Troyer2016}. A quantum emulator is a classical computer program that simulates a quantum processor on a classical computer. In comparison with the real quantum processors, quantum emulators are easily accessible and perfectly suite for quantum algorithms design and testing purposes.

In this report, we present one more exact state-vector quantum computer emulator. It is implemented in Rust programming language that makes it easily maintainable and less error-prone in comparison with C/C++ based implementations. Meanwhile, its speed and memory efficiency is at the same level with C/C++ based implementations. We performed and present here a wider range of benchmarks of our emulator ranging from Quantum Fourier transform to simulation of a quantum Ising model.

\section{Main features of the emulator}

Here we discuss main features of the presented emulator. 
The emulator consists of three modules. 
The first module is the kernel that is responsible for quantum computer emulation. 
The kernel is implemented from scratch without use of heavy numerical linear algebra packages. 
This allows one to achieve higher performance by narrowing the purpose of the kernel to a particular small set of tasks (one- and two-qubit operations). 
The kernel stores the state of a quantum processor in RAM exactly, i.e. for $n$ qubits it stores a contiguous array in memory with $2^n$ complex numbers in single- or double-precision floating-point format. 
It is capable of performing all typical operations over a state, i.e. apply a one- or two-qubit gate to certain qubits, measure a particular qubit in the standard basis, reset a particular qubit. 
When one applies a quantum gate to a state it performs update of a state in-place, i.e. it does no allocate any extra memory in a heap. 
This is an important point, since the main drawback of such kind of emulators is the enormous memory requirements that scales exponentially with number of qubits. 
It utilizes sparsity of gates when it is possible. 
For example, when one applies a \textsf{CPhase} gate to the state it only multiplies certain elements of the state by the corresponding phase factor, it does not perform a full tensor contraction that is necessary in general case. 
To fully utilize capabilities of modern CPUs, it also splits each operation into sub-operations which are executed within a thread pool in parallel. 
The current version of the kernel does not support accelerators such as GPUs or TPUs, although we are planning to add this option in future releases.

The second module implements support of OpenQASM 2.0 language \cite{cross2017open} for quantum circuits programming. 
Its main part is the virtual machine, which compiles OpenQASM 2.0 code into native instructions of the kernel and executes them. 
Thus, the presented emulator can be used as a backend for most of the modern quantum computing framework, since almost all of them support compilation of quantum circuits to the OpenQASM 2.0 format.

The first two modules are implemented entirely using the Rust programming language, which is one of the modern alternatives to C/C++ programming languages. 
Even though Rust programming language gets popular very fast, it is complicated programming language and not yet widely spread in the ``quantum'' community. 
To mitigate the complexity coming along with the Rust programming language, we have included the third module that wraps the Rust based API into Python API. 
Therefore, the entire emulator is user friendly and can be used by a user with basic knowledge of the Python programming language.

It is also worth to note, that the given emulator use minimal amount of external dependencies. 
As already has been said, it does not use heavy linear algebra packages that are redundantly complex for such an emulator.

\begin{figure}
    \centering
    \includegraphics[scale=0.8]{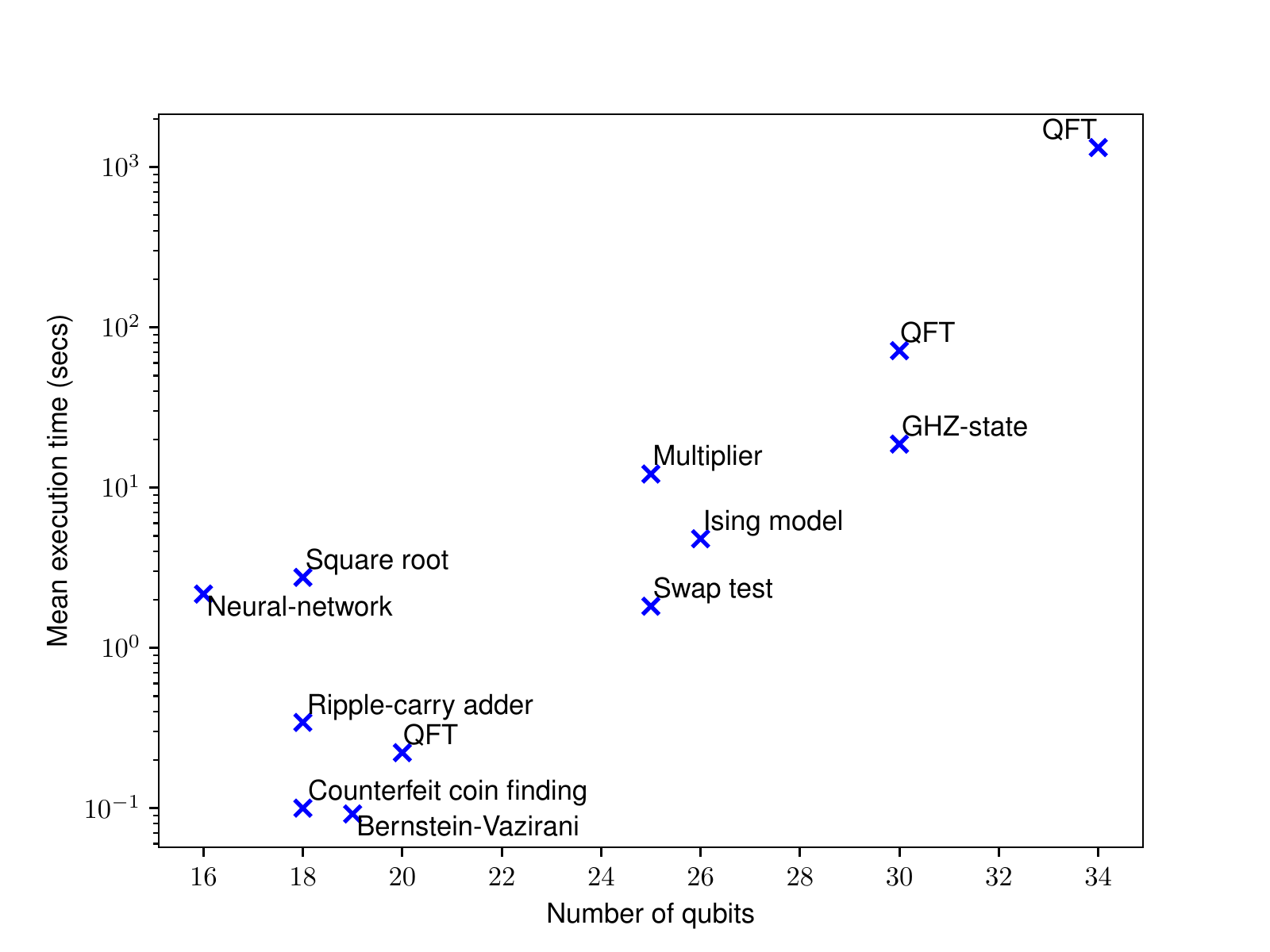}
    \caption{Comparison of algorithms in terms of number of qubits and execution time.}
    \label{fig:benchmarking}
\end{figure}

\section{Benchmarks}

For understanding capabilities of the developed emulator, we provide execution time statistics collected across multiple runs of different quantum algorithms. 
OpenQASM 2.0 code of all the quantum algorithms and their brief descriptions were taken from the GitHub repository of the paper~\cite{li2021qasmbench} (except the code for quantum Fourier transform for 20, 30, and 34 qubits). 
For each algorithm we performed a routine called a {\it benchmark}, i.e. each algorithm was run within the emulator multiple times, for each run one measured the execution time and calculated execution time statistics across runs: 
mean, maximal, minimal execution time and standard deviation. 

The following algorithms have been benchmarked and reported here: 
\begin{enumerate}
	\item Quantum Fourier transform (QFT) on 20, 30, and 34 qubits \cite{nielsen2002quantum}. 
	The entire circuits contained the layer of Hadamard gates necessary to prepare a superposition of all possible states, the QFT layer that drives the superposition state to the trivial one where all qubits in the state $\ket{0}$ 
	and a measurement layer that is necessary to validate the final state.
	\item
	The algorithm for quantum counterfeit coin problems \cite{iwama2012quantum} on 18 qubits. 
	The goal of the algorithm is to find all false coins in a set using a balance scale.  
	\item
	The quantum ripple-carry adder algorithm \cite{cuccaro2004new} on 18 qubits. It performs an addition of two binary numbers. 
	\item 
	Bernstein-Vazirani algorithm \cite{bernstein1993quantum} on 19 qubits. 
	The algorithm allows one to reconstruct a hidden bit-string defining a binary function $f(x) = s^\top x \ {\rm mod} \ 2$ in a single call of $f$. 
	\item
	The swap test \cite{barenco1997stabilization} on 25 qubits. This test aimed on measuring of how two quantum states differ. 
	\item 
	The quantum multiplier \cite{ruiz2017quantum} on 25 qubits. 
	The algorithm multiplies two binary numbers.
	 \item
	The example of a quantum neural network on 16 qubits. 
	\item
	The GHZ state preparation algorithm~\cite{greenberger1989going} on 30 qubits. 
	The algorithm drives the initial standard state (all qubits in $\ket{0}$ state) to the GHZ state. 
	\item
	The quantum algorithm that calculates the square root of a number \cite{babu2020quantum} on 18 qubits. 
	\item
	The quantum circuit that simulates dynamics of a quantum Ising model on 26 qubits. 
\end{enumerate}	
	
The summary of the benchmarks is given in Table~\ref{main_table}. 
We also provide a scatter plot in Fig.~\ref{fig:benchmarking} with algorithms comparison in terms of number of qubits and execution time. 
All the benchmarks were run on a single CPU machine with the following characteristics: CPU Intel(R) Core(TM) i9-10920X 3.50 GHz and 256GB of RAM. 
As it can be noted, the maximum number of qubits across the benchmarks is 34. 
This value is upper bounded by the available RAM: storing a quantum state of 34 qubits exactly with a single-precision floating-point format requires 128GB of RAM, 
storing a state of 35 qubits requires 256GB of RAM i.e. all the memory of a given machine that leads to a program crash. 

As one can see from the table, each benchmark is characterized by the number of parameters: 
the type of an algorithm; the number of qubits that were emulated; the number of independent runs of an algorithm; the number of gates that are native for the emulator; the number of measurements; the number of single-qubit resets. 
In contrast to a real quantum hardware, the complexity of a two-qubit gate implementation is roughly the same as the complexity of a one-qubit gate implementation. 
This is why we do not distinguish between them and calculate only the total number of gates.

To sum up the benchmarking results, one can note that the given emulator is capable to emulate an intermediate size quantum computer exactly on a small size computing server.

\begin{table}
\begin{flushleft}
\begin{tabular}{| p{3cm} | p{2.4cm} | l | l | l | p{2.2cm} | l | p{3.7cm} |} 
 \hline
Description & Execution time statistics (seconds) per run & Number of qubits & Number of runs & Number of gates & Number of measurements & Number of resets \\ 
 \hline\hline
 Quantum Fourier transform & \makecell[l]{mean: 0.222 \\ max: 0.467 \\ min: 0.128 \\ stddev: 0.072} & 20 & 100 & 230 & 20 & 0 \\
 \hline
 Quantum Fourier transform &\makecell[l]{mean: 71.54 \\ max: 71.70  \\ min: 71.46  \\ stddev: 0.07} & 30 & 10 & 495 & 30 & 0 \\ 
 \hline
 Quantum Fourier transform &\makecell[l]{mean: 1324.3 \\ max: 1326.7 \\ min: 1320.3 \\ stddev: 2.8} & 34 & 3 & 619 & 34 & 0  \\ 
 \hline
 Counterfeit coin finding problem &\makecell[l]{mean: 0.1 \\ max: 0.293 \\ min: 0.045 \\ stddev: 0.041} & 18 & 1000 & 71 & 18 & 0 \\ 
 \hline
 Quantum ripple-carry adder &\makecell[l]{mean: 0.343 \\ max: 0.769 \\ min: 0.155 \\ stddev: 0.122} & 18 & 1000 & 284 & 9 & 0 \\
 \hline
 Bernstein-Vazirani algorithm &\makecell[l]{mean: 0.092 \\ max: 0.233 \\ min: 0.061 \\ stddev: 0.023} & 19 & 1000 & 56 & 18 & 0 \\
 \hline

 Swap test to measure quantum state distance &\makecell[l]{mean: 1.821 \\ max: 1.832 \\ min: 1.815 \\ stddev: 0.005} & 25 & 10 & 254 & 1 & 0 \\
\hline
Quantum multiplier &\makecell[l]{mean: 12.16 \\ max: 12.21 \\ min: 12.14 \\ stddev: 0.02} & 25 & 10 & 1743 & 5 & 0 \\
\hline
Quantum neural network &\makecell[l]{mean: 2.168 \\ max: 2.948 \\ min: 1.214 \\ stddev: 0.32} & 16 & 100 & 2016 & 16 & 0 \\
\hline
GHZ state preparation &\makecell[l]{mean: 18.69 \\ max: 18.73 \\ min: 18.67 \\ stddev: 0.02} & 30 & 10 & 30 & 30 & 0 \\
\hline
Square root of a number &\makecell[l]{mean: 2.755 \\ max: 3.403 \\ min: 1.795 \\ stddev: 0.487} & 18 & 25 & 2300 & 13 & 65 \\
\hline
Ising model simulation &\makecell[l]{mean: 4.798\\ max: 4.806\\ min: 4.791\\ stddev: 0.004} & 26 & 10 & 280 & 26 & 0\\
\hline
\end{tabular}
\caption{Summary of the benchmarks.}
\label{main_table}
\end{flushleft}
\end{table}

\section{Code availability}

The current version of the emulator is 0.3.0. In sections below we describe main features of the current version of the emulator and present some benchmarking results.
The source code of the emulator and code for benchmarks is available under reasonable request.

\end{widetext}

\bibliography{bibliography.bib}

\newpage


\end{document}